# Why and How Coronavirus Has Evolved to Be Uniquely Contagious, with Uniquely Successful Stable Vaccines

Marcelo A. Moret (1), Gilney F. Zebende (2) and J. C. Phillips (3)

(1) SENAI CIMATEC Salvador BA Brazil

(2) Physics, Univ. Feira de Santana, BA Brazil

(3) Physics and Astronomy, Rutgers University, Piscataway

Abstract

Spike proteins, 1200 amino acids, are divided into two nearly equal parts, S1 and S2. We review here phase transition theory, implemented quantitatively by thermodynamic scaling. The theory explains the evolution of Coronavirus extremely high contagiousness caused by a few mutations from CoV2003 to CoV2019 identified among hundreds in S1. The theory previously predicted the unprecedented success of spike-based vaccines. Here we analyze impressive successes by McClellan et al., 2020, in stabilizing their original S2P vaccine to Hexapro. Hexapro has expanded the two proline mutations of S2P, 2017, to six combined proline mutations in S2. Their four new mutations are the result of surveying 100 possibilities in their detailed structure-based context   Our analysis, based on only sparse publicly available data, suggests new proline mutations could improve the Hexapro combination to Octapro or beyond.

**Introduction**

From static studies alone it is not clear why CoV-2 is so much more contagious than CoV-1, or why several more recent strains containing a few more mutations are even more contagious. Presumably this is caused by convergent evolution[1], which has not occurred recently in other airborne viruses like flu. Similarly, it is not clear why the spike vaccine has been so successful (~ 95%), compared to flu vaccines (~30-50%). A thermodynamic model based on phase transitions predicted the uniquely high success rate of spike-based vaccines in its abstract[2]. It also identified the mutations responsible for the unique asymptomatic first (prefusion) stage of infection that precedes structural rearrangements. Without new parameters the model also explained why only a few more mutations greatly enhanced viral spreading[3].

Most chemistry or physics graduate courses in statistical mechanics end with developments up to 1940. Probably the most important recent developments concern phase transitions in complex networks[4]. Many natural networks are self-organized and exhibit power-law dependence of numbers describing some quality or quantity. The power-law exponents are called fractals, and fractals always characterize second-order phase transitions more accurately than simple fractions[5]. Phase transitions occur because "some biological systems - parts, aspects, or groups of them - may extract important functional benefits from operating at the edge of instability, halfway between order and disorder, i.e., in the vicinity of the critical point of a phase transition. Criticality has been argued to provide biological systems with an optimal balance between robustness against perturbations and flexibility to adapt to changing conditions as well as to confer on them optimal computational capabilities, large dynamical repertoires, unparalleled sensitivity to stimuli, etc. Criticality, with its concomitant scale invariance, can be conjectured to emerge in living systems as the result of adaptive and evolutionary processes"[6]. "Biological regulatory networks are more distinguished from random networks by their criticality than by other macroscale network properties such as degree distribution, edge density, or fraction of activating conditions"[7].

Proteins are the most complex self-organized networks, so can we reasonably expect to learn much about them from studying simpler complex networks[4,5] which have a single property described by a single power law and fractal? No, but here there is a hidden simplifying single feature: all proteins have evolved in water, and function covered by one or two monolayers of water films. One can learn to quantify protein interactions with water films that shape protein transitions from resting states to functional states and return them reversibly to their resting states. This program was begun by biologists in the 1970's, when they began to develop hydropathicity scales. These scales measure how each of the 20 amino acids in proteins "like" (hydrophilic) or "dislike" (hydrophobic) water molecules. The idea proved so popular that by 2000 there were more than 100 biological hydropathicity scales; this has led chemists to conclude that all such scales have only qualitative value[8].

The next step is to use the very large 21st century database to study hydropathic curvatures of protein structures (as measured by solvent-accessible surface areas) of each of the 20 amino acids in turn. This required studying > 5000 segmental structures, and it produced a splendid result[9,10]. Instead of only one fractal, there are now 20 fractals, derived from power-law fits to segments of length L, with $9 \leq L \leq 35$. The existence of these fractals shows that many proteins have evolved in tandem to be on the edge of critical points associated with second-order phase transitions[6]. Their evolution has spanned very long times, whereas viruses have flourished over much shorter times. Thus most viruses (like flu or HIV) merely vary strains but show little proximity to criticality.

**Results**

The multiple successes of the phase transition model in describing the evolution of the S-1 part of spike proteins[2,3] raise an interesting question: can it also discuss the very impressive success of Hexapro vaccine mutants of the S-2 part[11] in achieving stability sufficient for production by growth in chicken eggs[12], where flu vaccines are also produced? Success was achieved through production and study of 100 variants of the previous S-2P mutant which involved two nearest neighbor Proline S-2 substitutions[12]. Selection of the 100 mutants was guided by common strategies used to stabilize similar fusion proteins. Thus the 100 mutants were distributed among 40 double Cys substitutions to form disulfide bonds, 31 cavity-filling substitutions, salt bridges, and only 14 prolines. Only prolines are used cooperatively in Hexapro; the other 86 "common" mutants all failed. This means that there is something special about CoV spikes. These are also the only viruses where the phase transition model has been successful, although it has been successful in describing the evolution of many proteins, even those involved in cytoskeletons[13,14]. The most active parts of Coronaviruses are their long spikes, which are immersed in water. Similarly proteins not only have evolved in water, but also their static shapes reflect dominant functionality through second-order phase transitions[2,3,6].

The 100 mutants were tested for their expression by an electrophoresis method that separates them by chain length, which had shown additional stability of prefusion S-2P against trimeric fusion[13]. The most stable 26 mutants (6 disulfide, 7 cavity-filling and salt bridged, 6 hydrogen bond, and 7 proline) were then tested further for monodispersity, thermostability, and quaternary structure. Thus expression alone had shown encouraging results for the six disulfide bonds, yet they (together with thirteen salt-bridging and cavity filling mutants), all failed the final test that combined the best-performing combined substitutions. "The expression of the Combo variants containing two disulfide bonds was generally half that of the single-disulfide variants, which suggests that they interfered with each other."[11]

Interference of coherent light wave amplitudes was first observed by Young in 1801; it can be either constructive or destructive. Here the disulfide bond interference is mostly destructive, as expected from a structure assembled by convergent evolution to be nearly optimal and coherent between domains spread over sliding thin film water windows of length W = 35 amino acids[2,14,15]. A single disulfide bond can accidentally enhance expression, but their combination works at cross purposes to disrupt the coherence of stabilizing water waves. The expressions of nine Combo disulfide bond variants were at most half of the disulfide bonds. Other experiments showed that even adding single disulfide bonds could cause expansion, not contraction of the spike globule, indicating partial disruption of coherent hydropathic compaction.

Aren't water waves only a new and merely arbitrary mechanism for explaining largely destructive interference for all spike substitutions except Proline? They are neither arbitrary nor new. The abstract of a paper discussing the action of the molecular motor dynein[14] included an explanation of a long-standing puzzle: "the energy used to alter the head binding and propel cargo along tubulin is supplied by ATP at a ring 1500 amino acids away … many details of this extremely distant interaction are explained by water waves quantified by thermodynamic scaling". Critics of water waves cited " heptad repeat regions that form coiled structures [~ 100 amino acids each] in the dynein stalks" as examples of "relatively long-range" hydropathic patterns, and did not discuss how the motor's energy could travel through 1500 amino acids to reach the binding head without coherent thin film water waves[16].

The two proline substitutions[13] (L976P and V977P) in the center (913-1032, Uniprot P0DTC2) of the longest helix of S-2 had already increased yield by a factor of 50, so further refinements were unlikely to be large. Some combinations of salt bridges and cavity-filling showed increased expression, but "multiple proline substitutions resulted in the most substantial increases in expression and stability". The analysis of 47 "combo" substitutions led to Hexapro, where the four best additional proline substitutions to S-2P (F817P, A892P, A899P, A942P)

"displayed higher expression than S-2P by a factor of 9.8, had a ~5°C increase in $T_m$, and retained the trimeric prefusion conformation."

Can phase-transition criticality suggest why these four (among 14 tested) proline substitutions combine so well? Note that three of these involve replacement of alanine (the second smallest amino acid) by proline, which differs from alanine mainly because its connectivity to the peptide backbone has increased from 1 to 2; proline's double connection is unique. This tightens and strengthens the fold with little change in shape or volume. The increase in local stiffness may reach no further than second neighbors, and does not disturb more distant packings, which have been optimized by evolution that brought CoV-1 closer to the critical point of CoV-2 (mainly through mutations of S-1)[2,3]. It would be interesting to see if bond network counting could provide additional insights into protein stiffening[17].

These broad general principles can be regarded as complementary to the extremely detailed structural analysis used to select the initial 100 mutants that have been studied so far[11,13]. Specifically they explain why combined proline substitutions have worked so well, and common alternatives have failed, to stabilize Coronavirus cooperatively. Note that dominance of peptide chain stiffening is consistent with not only with the triad of small alanine replacements, but also with F817P; F (phenylanine) is treated as cavity-filling, because it is a large neutral amino acid.

Further improvements in CoV vaccine are unlikely to be large, but are nevertheless important to vaccine production[12]. The present model can be tested by asking it to predict two more stiffening mutations that would improve Hexapro to Octapro. The 40 disulfide bonds previously studied covered the region 570 – 1117, and extended past the furin S-1/S-2 cleavage site at 685 (nearly all of S-2 and part of S-1). The 14 proline substitutions covered a shorter region 817-978. Within this region there are still many possibilities, but one can also look for two more proline sites outside this region, one below 817, and one above 978.

By far the most successful additional Hexapro replacement was A942P, near the start of a long helical region 913-1032, between two turns and near the center of Heptad Repeat 1 (920-970). One can find shorter helical regions listed on Uniprot P0DTC2, and from these find two candidates for Octapro, A766P (helical region 747-782) and A1174P (helical region 1142-1201). The second region contains Heptad Repeat 2 and was deleted without improvements,

6while if it is retained with A1174P, it could provide additional stiffening. Other possibilities are A871, 876, 879P, and inside the long helical region 913-1032, A(930,955, 972,989,1015,1016)P. Altogether, the critical phase transition model unifies the highlights of both S-1 and S-2 data, and provides a broad consolidation of extensive and still growing spike data.

Another unique feature of Coronavirus is the discovery of a continuum of conformation states of the CoV-2 spike, instead of well defined, stable spike conformations[18]. This is indicative of extreme dynamical balance between stability and adaptability characteristic of naturally evolved proteins[19] close to critical points. Similar (but much smaller) conformational continua must be present in many proteins as a consequence of coherent water wave fluctuations. These elastic deformations are connected to the resonant hydrophilic edge leveling that was previously used to explain the enhanced contagiousness of CoV-2 relative to CoV-1[2], and the even larger contagiousness of more recent strains (such as UK and Calif.)[3]. It may have important practical implications in the search for new antibodies. It is difficult to control such a range of conformations with a single antibody, while two antibodies could span a wide range. The most successful Coronavirus antibody so far is such a "cocktail" (casirivimab with imdevimab).

The general theory of evolutionary approach to criticality[6] suggests that wave patterns can be found both at the molecular and the cellular level. Membrane waves[20-22] are now proposed as important to immune signaling, with the waves passing through finger-like membrane protrusions of cells[23]. These microvilli are cellular analogs of hydrophilic extrema in proteins. They exhibit anomalous diffusion and fractal organization[24]. Two treatments can interfere destructively[22].